\documentclass{emulateapj}

\usepackage{color}
\def\teff{\mbox{T$_{\rm eff}$}}
\def\logg{\mbox{log~{\it g}}}
\def\vmicro{\mbox{$\xi_{\rm t}$}}
\def\kmsec{\mbox{km~s$^{\rm -1}$}}

\shorttitle{Na and O in NGC 6791}
\shortauthors{Bragaglia et al.}

\begin{document}

\title{Searching for Chemical Signatures of Multiple Stellar Populations 
in the Old, Massive Open Cluster NGC~6791.}

\author{
Angela Bragaglia\altaffilmark{1},
Christopher Sneden\altaffilmark{2,3,4}
Eugenio Carretta\altaffilmark{1},
Raffaele G. Gratton\altaffilmark{5},
Sara Lucatello\altaffilmark{5},
Peter F. Bernath\altaffilmark{6},
James S. A. Brooke\altaffilmark{7}, and
Ram S. Ram\altaffilmark{7}
}

\altaffiltext{1}{INAF-Osservatorio Astronomico di Bologna, 
via Ranzani 1, 40127 Bologna, Italy; angela.bragaglia@oabo.inaf.it, 
eugenio.carretta@oabo.inaf.it} 

\altaffiltext{2}{Department of Astronomy and McDonald Observatory, C1400,
University of Texas, Austin, TX 78712, USA; chris@verdi.as.utexas.edu}

\altaffiltext{3}{Department of Astronomy and Space Sciences,
                 Ege University, 35100 Bornova, Izmir, Turkey}

\altaffiltext{5}{Visiting Astronomer, Kitt Peak National Observatory.
The WIYN Observatory is a joint facility of the University of
Wisconsin-Madison, Indiana University, Yale University, and the National 
Optical Astronomy Observatory.
This research has made use of the Keck Observatory Archive (KOA), which is
operated by the W. M. Keck Observatory and the NASA Exoplanet Science Institute
(NExScI), under contract with the National Aeronautics and Space
Administration.}

\altaffiltext{4}{INAF-Osservatorio Astronomico di Padova, vicolo 
dell'Osservatorio 5, 35122 Padova, Italy; raffaele.gratton@oapd.inaf.it, 
sara.lucatello@oapd.inaf.it}

\altaffiltext{6}{Department of Chemistry \& Biochemistry, Old Dominion 
                 University, 4541 Hampton Boulevard, Norfolk, VA 
                 23529-0126, USA, pbernath@odu.edu}

\altaffiltext{7}{Department of Chemistry, University of York, Heslington, 
                York, YO10 5DD, UK; jsabrooke@gmail.com, rr662@york.ac.uk}

\begin{abstract}
Galactic open and globular clusters (OCs, GCs) appear to 
inhabit separate regions of the age-mass plane. 
However, the transition between them is not easily defined because there is 
some overlap between  high-mass, old OCs and low-mass, young GCs.  
We are exploring the possibility of a clear-cut  separation between
OCs and GCs  using an abundance feature that has  been found so far 
only in GCs: (anti)correlations between light elements.  
Among the coupled abundance trends, the Na-O anticorrelation
is the most widely studied.
These anticorrelations are the signature of
self-enrichment, i.e., of a  formation mechanism that implies multiple
generations of stars.  Here we concentrate on the old, massive, metal-rich OC
NGC~6791.    We analyzed archival Keck/HIRES spectra of 15 NGC~6791 main
sequence  turn-off and evolved stars, concentrating on the derivation of C, N,
O,  and Na abundances.
We also used WIYN/Hydra spectra of 21 evolved stars (one is in common).
Given the spectral complexity of the very metal-rich NGC~6791 stars, we
employed spectrum synthesis to measure most of the abundances.  We confirmed the
cluster super-solar metallicity and abundances of Ca and Ni that have been
derived in past studies. More importantly, we did not detect any significant
star-to-star abundance  dispersion in C, N, O and Na.  Based on the absence of a
clear Na-O anticorrelation, NGC~6791 can still  be considered a true OC, 
hosting a single generation of stars, and not a low-mass GC.

\end{abstract}

\keywords{globular clusters: general --- open clusters and associations: 
general --- open clusters and associations: individual (NGC 6791) ---  
Stars: abundances}

\section{INTRODUCTION\label{intro}}

The separation between Galactic open and globular clusters (OCs, GCs) is 
ambiguous, and probably depends on mass, age, and Galactic population 
membership.   
Photometric data are insufficient to address this question; the 
color-magnitude diagrams (CMDs) of the oldest open clusters are morphologically 
similar to those of metal-rich globulars.  
Fortunately, the problem can be attacked in a different way, by 
searching for some  chemical signatures that are unique to one
or the other cluster type.
Almost all Galactic GCs studied to date with
sufficient sample sizes exhibit star-to-star variations in the abundances of 
light elements C, N, O, Na, Mg, and Al, while their heavy element abundances
usually are very uniform  \cite[see the reviews by][and references
therein]{araa04,rev12}.   

The light-element variations in GCs are not random, but occur in distinct  
patterns.   
Particularly striking is the observed Na-O anticorrelation discovered by 
the Lick-Texas group (see the review by \citealt{kraft}).  
This feature has been found in (almost) all clusters studied so far  
\citep[e.g.][and references therein]{carretta09,carretta10},  even in 
the low-mass ones. 
Possible exceptions are Ter~7, Pal~12 \citep{tautvaisiene,sbordone,cohen},  
where however only five and four stars were analyzed, respectively, 
and Rup~106 \citep{vil13}. 
This anticorrelation is noticeably absent in field halo stars 
\citep[e.g.,][]{gra00}, while it is found at all evolutionary phases in GCs
\citep[e.g.,][]{gratton01,cm05}.  This phenomenon is now firmly  attributed to
the pollution  of a previous generation of cluster stars, and is clearly
dependent on the  formation and early evolution of clusters.

The Na-O anticorrelation is such a distinctive characteristic of GCs  that
\cite{carretta10} proposed to use it as a definition  of GCs:  these are
clusters able to sustain self-pollution and to  produce a second
generation\footnote{The model proposed by \citet{bastian}, based on accretion on
low-mass pre-main-sequence stars from massive stars, does not actually require
multiple generations, but a large cluster mass is still needed.} of stars with
modified composition (enriched in  Na and depleted in O). The Na-O
anticorrelation has been amply studied in Milky Way GCs: see e.g., the work of
the Lick-Texas group (reviewed by \citealt{kraft}  and \citealt{araa04}); the
large, homogeneous sample of GCs observed  with FLAMES at VLT (\citealt{carretta09}
and references therein);  or the recent work on M~13 by \cite{jp12}.  This
anticorrelation has also been found in GCs of the Large  Magellanic Cloud
(\citealt{johnson06,mucciarelli}) and the Fornax  dwarf galaxy
(\citealt{letarte}). However, light element abundances have almost never been
investigated in OCs with the same kind of large samples.   A summary of O and Na
results for OC's can be found in  \cite{desilva},  who collected literature data
from various sources and normalized them  to a common abundance scale.  They
found no evidence for internal (single cluster) dispersion in either  element,
but cautioned about the paucity and the heterogeneity of the  published
samples. 

Although most OCs are far less massive than GCs, there is a ''grey area" 
containing low-mass and/or young GCs and high-mass and old OCs \cite[][see in
particular their Figure~2]{carretta10}.  We are studying clusters in this area
of the mass-age plane.  We have already presented results on 30 stars of  the
old, massive OC Berkeley~39 \citep{be39}. Analyses of our FLAMES/VLT spectra
indicated no star-to-star  spread in any of the measured element, in particular
in Na and O. Recently, we have investigated Ter~8, a low-mass,  low metallicity
([Fe/H]~$\simeq$~$-$2.3) GC belonging to the Sagittarius  dwarf spheroidal
(\citealt{car13}).  In our sample, one star out of the 20 observed red giants
has an overabundance of Na. This indicates the presence of a very small fraction
of second-generation  stars, keeping Ter~8 in the GC category as defined by
\cite{carretta10}.

We targeted also another interesting object, \object{NGC 6791}.  
This OC is very old (about 9~Gyr, e.g., \citealt{king05}, or about 7-8 Gyr, 
\citealt{brogaard12}), metal-rich ([Fe/H]=0.30-0.45, e.g.,  
\citealt{pg98,gra6791,boes09,occam}), and massive ($\sim10^4 ~M_\odot$,   e.g.
\citealt{liebert}, \citealt{platais11}). 
To account for the combination of age, mass, 
and a metallicity much higher than that of the other clusters at similar 
Galactocentric positions, recent studies have proposed hypotheses on the 
formation site of NGC~6791, 
\citep[see e.g.][]{jilkova} for a recent calculation of its orbit. 

Given all these characteristics, NGC~6791 is an important link between the OCs
and GCs.  To have an idea of the position of
NGC~6791 in the mass-age plane,  see Figure~1 in \cite{be39}, where $M_V$ is
used as a proxy for mass  (and note that its $M_V$ has been revised upward by
\citealt{buzzoni}). 
NGC~6791 is an ideal target to test the lower mass limit for the Na-O 
anticorrelation (i.e., of multiple generations of stars), because of its 
relatively large age and mass.  
Indeed, while we were revising the analysis of
our data, \cite{geisler} presented evidence of a Na-O anticorrelation, on
the basis of spectra obtained with the HIRES spectrograph on the Keck I
telescope (hereafter Keck/HIRES) and the Hydra multi-object spectrograph on the
WIYN telescope (WIYN/Hydra), speculating on the nature and origin  of this
cluster.

We present here a light-element abundance study of archival Keck/HIRES  spectra
of 15 main-sequence turn-off (MSTO), subgiant branch (SGB),  and red giant
branch (RGB) members of NGC~6791. These stars are listed in
Table~\ref{tab-hires}. We supplement this work with an investigation of Na
abundances from a  sample of 21 RGB and red clump (RC) stars obtained with
WIYN/Hydra (one of which in common with the Keck sample), which are listed in
Table~\ref{tab-hydra}.
The HIRES data are discussed in \S\ref{keck}, with subsections that describe the
targets, the observations and reductions, atmospheric parameters, and abundance
results. Then in \S\ref{hydra} we consider the Hydra data in a similar manner.
Our results from both data sets are merged and compared in \S\ref{paperend}, 
and a summary is presented in \S\ref{summary}.

\section{NGC 6791 LIGHT ELEMENT ABUNDANCES FROM HIRES SPECTRA\label{keck}}

NGC~6791 presents two formidable challenges for medium-high  resolution
spectroscopic analyses:  it is fairly distant, about 4~kpc, and it  is very
metal-rich, [Fe/H]~$\simeq$ $+$0.4 dec, as discussed in \S\ref{intro}. The large
distance of this cluster makes even its most luminous stars  faint observational
targets: $V$(RGB~tip)~$\simeq$~14. Only three of the stars in the HIRES sample
are brighter than  $V$~$\simeq$ 17.1, so all of the HIRES data have
$S/N$~$\lesssim$ 50. The high metallicity ensures that the RGB stars, which are
the brightest NGC~6791 members, have mostly strong lines that are sensitive to 
microturbulent velocity choices. The spectral features often are also blended,
and it is difficult to find many clean lines with which to perform a traditional
abundance analysis. Additionally only Fe is represented by both neutral and
ionized species transitions  in the spectral region usable for all the HIRES
spectra (about 5500$-$6500~\AA). These cautions should be remembered when
considering the abundance results presented here.

\subsection{The Stellar Sample\label{keckspec}}
                                                              
We queried the Keck Observatory Archive (KOA)\footnote{
https://koa.ipac.caltech.edu/cgi-bin/KOA/nph-KOAlogin} and found spectra of 19
stars observed with Keck/HIRES (configured to have resolving power $R \equiv
\lambda/\Delta\lambda \simeq$ 45,000).   The spectra had been obtained
from 1999 to 2009 under several observing programs by three Principal
Investigators (PIs): Ann Merchant Boesgaard (10 stars), Fabio Bresolin (six
stars), and Geoffrey W. Marcy (three stars).  The spectra were acquired using
different setups, but they all included  the wavelength region containing the
same Na and O lines present   in the Hydra spectra. To our best knowledge,
analyses have been published only for two stars in \citet{boes09} and five
stars in \cite{geisler}.

In Table~\ref{tab-hires} we list
coordinates and photometric data from \cite{stetson} and 2MASS for these stars;
they will be called by their numbers  given in  \cite{stetson}. We show
their evolutionary phases in Figure~\ref{NEWcmd}. 
About half of the stars are at
or near the main sequence turn-off (MSTO),  six are located at the base of the
RGB, and three are on the RGB  just below the RC. Since we have
a large enough sample, we preferred to exclude a few dubious cases.
We discarded as non members
or possible non members (see Table~\ref{tab-hires} and Figure~\ref{NEWcmd}) a few stars that showed metallicity discordance with the main body of the sample
or different RV (star 2130; note that we did not correct the absolute RV zero point of the HIRES reduced spectra, so we worked differentially).

We retrieved from KOA the raw data and the extracted/calibrated spectra and used
the second for our analysis. The extracted spectra's quality was
checked on two extreme cases, star 5744  (at the main sequence turn-off,
observed in 1999 with the original single chip setup)  and star 8351 (on the
RGB, just below the RC, obseved with the 3-chips setup),  i.e., one of the
faintest, one of the brightest objects. We retrieved four and three exposures,
respectively, and the relevant calibration files. We used MAKEE, the pipeline
reduction for HIRES and proceeded with the analysis as for the  spectra
presented in the paper (see next sections).  We found that the original and
newly reduced spectra are essentially identical when overplotted.  
                  
We examined the individual exposures of each star (two to three for  the bright RGB stars, and up to seven for the MSTO ones), overplotting them to be
sure of their correct wavelength registration. We excised the noise spikes in
each spectrum before.
Finally, we averaged the individual
spectra  performing a straight mean, since the individual exposures for
each star were comparable to each other, as deduced from the S/N information in the Keck archive.

Even with these co-additions, the resulting noise levels in the spectra
are large. We estimated $S/N$ ratios in the 6100$-$6200~\AA\ spectral region in
two ways. First, we searched for small spectral intervals that appeared to have
only very weak or absent absorption lines and directly measured the noise
levels. Second, we compared synthetic spectra computed for abundance
determinations with the spectra for individual stars (see below), and took the
standard deviation $\sigma$ of the residual values as a second indicator of spectrum noise. The means of
these two estimates, which were always similar, are given in 
Table~\ref{tab-hires}. All of the HIRES spectra have $S/N$~$<$~50.
 Our values are generally smaller than those published in \citet{boes09,geisler}. This is probably due to different ways of estimating
$S/N$ and for star 7649, to the use of only two of the four spectra.

\subsection{Model Atmosphere Parameters and Metallicities\label{keckmods}}

NGC~6791 has been the subject of several general abundance  studies; see e.g.,
the sources cited in \S\ref{intro} and references therein. Our chief interest is
in derivation of light element abundances, and therefore we did not undertake a
new comprehensive model atmosphere analysis. Effective temperatures \teff\ and
surface gravities \logg\ were derived from the cluster photometry.

We adopted $V$ magnitudes, and $B-V$, $V-I_C$ colors, and the star numbering
system from \cite{stetson}. These were input to the relations of \cite{rm05}
(who use $V-I_C$) to derive two temperature  estimates, which were then averaged
to produce the \teff's given in  Table~\ref{tab-hires}.   Bolometric corrections
were computed with relations given by \cite{alonso}. For NGC~6791 we assumed a
distance modulus $(m-M)_V=13.45$, and a  reddening E$(B-V)=0.15$, as we did in
\citet{gra6791}. 
Differential reddening is a rather common occurrence in OCs. NGC~6791 is not an
exception \citep{platais11,brogaard11},  but the magnitude of its differential
reddening is small. \citet{brogaard11} published their photometry and
differential reddening corrections, from which we could identify the targets.
Figure~\ref{NEWdiffred} shows that the estimated correction is never large for
the cluster and is confined to $\pm0.02$ mag for our targets.  Since we have not
applied this correction, our temperatures might be wrong   by about 20-30 K for
an individual star. This would imply differences in the derived abundances of
less than about 0.01 dex.

We also assumed a mass of $M$~=~1.1~$M_\odot$ for all stars.  
\citet{brogaard12} determined the mass of stars on the lower RGB of NGC~6791
using several eclipsing binary systems; they found $1.15\pm0.02~M_\odot$, in
very good agreement with our choice.   NGC~6791 was observed by the $Kepler$
satellite and masses were derived  from asteroseismology of evolved stars 
\citep[$1.20\pm0.01~M_\odot$,][]{kepler}.  These values are all consistent,
considering the uncertainties and the  systematics involved in the derivation of
masses through photometry and  stellar evolutionary models, eclipsing binary
systems, and asteroseismology.  The fact that we adopted a single mass value for
all our targets is  not critical for the analysis; stars at the MSTO and on the
RGB have very similar masses in a cluster about 8 Gyr old. Furthermore,
\citet{miglio12} found from $Kepler$ data that stars on the RC and the RGB have
only a small difference in mass (about 0.1 M$_\odot$), further supporting our
choice of a single mass value for all stars.  The average $Kepler$ gravities are
larger than ours by about 0.07 dex \citep[see, e.g.][and private communication
by Miglio]{stello}, but this offset has a completely negligible effect on the
abundance derivations.
                                                              
The temperatures, magnitudes, reddenings, bolometric corrections, distance 
modulus, and mass were then combined in the standard $L-\teff-M$ 
relationship to yield the gravities listed in Table~\ref{tab-hires}.
Model atmospheres with these parameters were generated by interpolation
in the \cite{kur09}\footnote{ http://kurucz.harvard.edu/} grid, using
software developed by Andy McWilliam and Inese Ivans.

As a check on the model atmospheres, we measured some  lines of \ion{Ca}{1},
\ion{Fe}{1}, \ion{Fe}{2}, and  \ion{Ni}{1} in the list of \cite{gra03} on the
HIRES spectra  (see \citealt{gra6791} for the $gf$ values of iron lines), and derived  abundances of these elements for each star. The line
parameters given by \citeauthor{gra03} were  adopted without change in order to
be consistent with previous cluster  abundance studies by our group. We used
software developed by \cite{fit87} to measure equivalent widths, and used the
line analysis and synthetic spectrum package MOOG  (\citealt{sne73}).\footnote{
Available at http://www.as.utexas.edu/~chris/moog.html.} Almost all of the
measured lines are saturated to some extent in the very metal-rich NGC~6791
cluster, and we have no independent way to assess microturbulent velocities.
Therefore we adopted the gravity-dependent microturbulent  velocity relation
recommended by \cite{gra96}:  
\vmicro~= $-0.322\times\logg\ + 2.22$.
The \vmicro\ values are entered into Table~\ref{tab-hires}. For the three RGB
stars \vmicro~$\simeq$~1.4~\kmsec, while for all of the higher gravity stars
\vmicro~$\simeq$~1.0~\kmsec. See \cite{gra6791} for a more detailed discussion
of the  microturbulence issue in NGC~6791.

With our adopted values of \teff, \logg, and \vmicro, we derived consistent Fe
abundances for all 15 stars of the Keck sample. Averaging the abundances of
\ion{Fe}{1} and \ion{Fe}{2}, we found $<$log~$\epsilon$(Fe)$>$~=~7.88
($\sigma$~=~0.06, 6$-$7 lines in each star) for the Keck sample, or
[Fe/H]~=~$+$0.37 dex, using the solar Fe abundance derived by \cite{gra03},
 i.e. 7.54. For the
other elements we derived: $<$log~$\epsilon$(Ca)$>$~=~6.59 ($\sigma$~=~0.12,
5$-$6 lines) or [Ca/H]~=~$+$0.32; and $<$log~$\epsilon$(Ni)$>$~=~6.73
($\sigma$~=~0.07, 5$-$6 lines) or [Ni/H]~=~$+$0.45. Our [Fe/H] value is 0.10
lower than that derived by \cite{car6791}, but  our [Ca/H] and [Ni/H] values are
very consistent with the earlier study. This is satisfactory agreement, given
that the lines employed here are  not optimal for the very strong-lined spectra
of NGC~6791 stars.

\subsection{Abundances of C, N, O, and Na\label{keckcnona}}

Spectral features of C$_2$, CN, and [\ion{O}{1}] were used  to derive the CNO
abundances of the NGC~6791 HIRES sample. Details on the transitions will be
given in subsequent paragraphs. Molecular CO formation ties the C, N, and O
abundances  together in cool stars, and one must derive their abundances as a
group, especially for the RGB stars. Additionally, it is necessary to estimate
the CN contamination  of the [\ion{O}{1}] feature prior to extracting an O
abundance.  Therefore we iteratively derived the CNO abundances in the following
manner. First, an initial [O/Fe]~=~0 was assumed, and a preliminary C abundance
was estimated from C$_2$ features. Then a preliminary N abundance was set from
the CN strenghts in the  6290$-$6310~\AA\ region. A new O abundance was derived
from the 6300.3~\AA\ line. The process was repeated, this time expanding the
number of the CN features to include several more spectral regions, until
satisfactory consistency  was achieved between C$_2$, CN, and [\ion{O}{1}]
features.

{\textit{Carbon:} 
We derived C abundances from the C$_2$  Swan system (${\rm d}^3\Pi - {\rm
a}^3\Pi$) 0-1 bandhead near 5635~\AA. The C$_2$ bands are very weak but always
detectable on our spectra. The synthetic spectrum line lists were generated
beginning with new laboratory line data from \cite{bro13a} and 
 \cite{ram14}.
To the C$_2$ lines we added atomic lines from the 
\cite{kur09}\footnote{http://kurucz.harvard.edu/}  compendium. Empirical
corrections to the transition probabilities of the atomic lines were then made
through synthetic/observed spectrum matches to the spectrum of Arcturus. We
adopted a model atmosphere for this star with parameters recommended by
\cite{pet93}:  \teff~=~4300~K, \logg~=~1.50, \vmicro~=~1.7~\kmsec, and
[M/H]~=~$-$0.5 dex. With this model and the line list we generated synthetic 
spectra which then were convolved with a Gaussian smoothing function and matched
to the electronic version of the Arcturus spectral atlas
\citep{hin00}.\footnote{www.noao.edu/archives.html} Alterations to the log~$gf$
values in the line list were made until satisfactory observed/synthesized
matches were obtained.

We then applied this line list and the NGC~6791 model atmospheres  described
above to generate synthetic spectra, which were smoothed  with Gaussian
functions to match the broadening of the  NGC~6791  HIRES data. The C abundances
were estimated from synthetic/observed spectrum matches. We show C$_2$ bandhead
syntheses and observations for  two stars in Figure~\ref{c2synsc}. These stars
have been chosen for their contrasting CMD positions. Panel (a) shows
the SGB star 15592 (\teff~=~5675~K, \logg~=~4.10), and panel (b) shows the RGB
star 7347 (\teff~=~4445~K, \logg~=~2.55). The HIRES spectra  are of modest $S/N$
but the C$_2$ bandheads clearly are present in these and all but one other
program star. We did not attempt to derive a C abundance for star 5796, as its
spectrum is of very low quality near 5630~\AA. For the rest of the sample, the
final C abundances  are given in Table~\ref{tab-abund}. We have previously
cautioned about the influence of CO formation on abundances of C and O. However,
CO formation is small for the eight stars with \teff~$>$~5000~K. As an example,
molecular equilibrium computations  done for star 5744 (\teff~=~5365~K) as part
of the CNO abundance  computations indicate number density ratios that  $n$(C or
O)/$n$(CO)~$>$ 50 in line-forming regions ($\tau$~$\sim$ 0.5). These ratios are
even larger for the warmer SGB and MSTO stars of NGC~6791; C is completely
decoupled from O. Discussion of C in the RGB stars will be given below.

\textit{Nitrogen:} 
We attempted to determine N abundances in all program stars. Absorption features
of the CN red system (${\rm A}^2\Pi - {\rm X}^2\Sigma^+$)  occur throughout the
red-infrared spectral domain  ($\lambda$~$\gtrsim$~6000~\AA) of NGC~6791 stars.
A new laboratory study by \cite{bro13b} gives accurate wavelengths,  excitation
energies, and transition probabilities for the $^{12}$CN  lines, and 
 similar data for $^{13}$CN and C$^{15}$N are
provided by \cite{sne14}.
Beginning with these sources,
 we constructed atomic and molecular line lists  for CN in the
manner described above for C$_2$.  We examined synthetic/observed spectrum
matches of CN-dominated features  in three spectral regions:  6100$-$6200~\AA\
and 6290$-$6310~\AA\  in all stars, and 7900$-$7980~\AA\ in six of the coolest
program stars. CN absorption is much stronger  in the redder spectral
region, thus yielding the more reliable N abundances. Unfortunately, no HIRES
spectra of this longest-wavelength  domain are available in the KOA archives for
warmer NGC~6791 stars.

In Figure~\ref{cnsyns} we show small portions of the spectrum near 6300 and
7910~\AA\ of  the RGB star 7347. The N abundances have been altered by large
amounts, 0.6~dex, in the three displayed syntheses in order to show clearly the
spectral features that are dominated by CN. Inspection of the two figure panels
reveals that the CN line strengths are about a factor of two larger in the
redder spectral region; CN absorption dominates that from all other species in
this domain. We estimated N abundances in multiple small wavelength intervals of
the spectral regions named above, and averaged the result to give the N 
abundances reported in Table~\ref{tab-abund}. Happily, the CN features in all
spectral regions yield consistent  N abundances, with scatter $\lesssim$0.05~dex
from the various estimates that were made. Unfortunately we cannot report N
abundances for any NGC~6791 star with \teff~$>$~5000~K; we have no 7900~\AA\
spectra for any of them, and the CN strengths in their 6100$-$6300~\AA\ spectra
are too weak to produce reliable N abundances.

\textit{$^{12}$C/$^{13}$C:}
For all but one star, the HIRES wavelength coverage does  not extend beyond
7990~\AA, thus missing the $^{13}$CN transitions that have been the favorite
features for decades in deriving carbon isotopic ratios of cool stars (e.g.,
\citealt{day73}). The RGB star 9609 does have a spectral order containing
8030$-$8090~\AA. We repeated the synthetic/observed spectrum matches in this
region for 9609, this time accepting the CNO abundances from other spectral
regions but varying the assumed $^{12}$C/$^{13}$C. The $^{13}$CN absorption was
very weak, yielding only rough isotopic ratios: our analyses suggest that
$^{12}$C/$^{13}$C~$>$~15. We were able to rule out very low carbon isotopic
ratios in this star. Additionally, we found no evidence for strong $^{13}$CN
absorption in the 7900~\AA\ spectra of other NGC~6791 RGB stars, although its
spectral features are all blended in this region. 
Typical globular cluster stars exhibit 
 $^{12}$C/$^{13}$C~~$\simeq$~3$-$10
(e.g., \citealt{bro89}, \citealt{sun91},
\citealt{bri94}, \citealt{she03},  \citealt{pav03}, \citealt{car05},
\citealt{val11}). We did not detect large amounts of $^{13}$C, but
definitive statements on the $^{12}$C/$^{13}$C must await studies specifically
targeted to obtain spectra of the best $^{13}$CN features.

\textit{Oxygen:} 
We derived O abundances from the [\ion{O}{1}] 6300.3~\AA\ line. This transition
is considered to be the most reliable O abundance indicator in cool stars, but
it is not a simple spectral feature. In addition to the [\ion{O}{1}] line at
6300.31~\AA, there is a \ion{Ni}{1} line at 6300.34~\AA, and a $^{12}$CN red
system  line at 6300.29~\AA. The \ion{Ni}{1} has been well-documented (e.g.,
\citealt{all01},  \citealt{caf08}), and its transition probability has been
derived  from a laboratory study (\citealt{joh03}). The CN line (rotational
transition R$_1$(14.5) of the  10$-$5 vibrational band) has been considered far
less.   It is tabulated in many CN lists (e.g., in the  \citealt{kur09}
database) but neither its wavelength nor its oscillator  strength has been
secure in the past. There are many CN red system transitions near 6300~\AA\ as
discussed below, but only this particular line can potentially compromise the 
[\ion{O}{1}] feature. Here we used the parameters for this line (wavelength,
excitation  potential transition probability) given in the new extensive
laboratory study by \cite{bro13b}, and a CN dissociation energy D$_0$~=7.6~eV 
(this value is still uncertain by $\simeq\pm$0.1~eV).

We performed tests with repeated syntheses of  the RGB star 7347  to assess the influence
of [\ion{O}{1}], \ion{Ni}{1}, and CN on the  6300.3~\AA\ feature in the cooler
program stars. The synthetic absorption line that approximates the observed line
has EW~$\simeq$~46~m\AA. Using the final abundances for this star, and
synthesizing the 6300.3~\AA\ components separately, we find
EW([\ion{O}{1}])~$\simeq$~38~m\AA,  EW(\ion{Ni}{1})~$\simeq$~12~m\AA, and
EW(CN)~$\simeq$~5~m\AA. Incipient saturation of the total feature makes its  EW
less than the sum of its three individual components. Thus the CN contamination
is somewhat more than 10\%, but the \ion{Ni}{1} contamination is about 30\%. The
influence of the Ni transition at 6300.3~\AA\ cannot be neglected  even in the
RGB spectra of NGC~6791.

In Table~\ref{tab-abund} we give the final O abundance  estimates for the HIRES
RGB sample. The [\ion{O}{1}] line are very weak in the warmer MSTO and SGB
stars, and on average the spectroscopic $S/N$ values of these fainter stars are
too low to permit reliable O abundance determinations,  so the reader should take also the measured values with caution.

\textit{Sodium:}
We derived Na abundances from the \ion{Na}{1}  6154.23, 6160.75~\AA\ doublet via
spectrum synthesis, due to the general line crowding surrounding these otherwise
simple transitions. The mean Na abundances from these lines are entered in
Table~\ref{tab-abund}.   
In panels (a) and (b) of Figure~\ref{nacasyns} we show 
synthetic and observed spectra of the two \ion{Na}{1} lines and  
several \ion{Ca}{1} lines in the NGC~6791 subgiant star 15592, and in 
panels (c) and (d) we show these spectral features in the RGB star 7347.
Inspection of this figure makes it clear that while the cooler
RGB star has a much stronger-lined spectrum than does the warmer SGB star, 
the \ion{Na}{1} and \ion{Ca}{1} features are saturated in both stars.
Both Na transitions also have small amounts of blending by lines of 
other species, particularly the 6160~\AA\ line which has two strong  Ca 
features to the red of it.
However, our derived
abundances from the Na lines are in accord:
$<$log~$\epsilon$(6160)~$-$~log~$\epsilon$(6154)$>$ = $+$0.02 
($\sigma$~=~0.09). The fairly large star-to-star scatter in this difference
emphasizes the utility in deriving Na abundances from both lines.

The neutral Na species is subject to departures from LTE; see, e.g., \cite{lind}
for a recent evaluation. Had we only stars of similar evolutionary status in our
sample, we could avoid corrections for NLTE entirely and work only
differentially. For the 6154, 6160~\AA ~doublet the NLTE corrections are quite
small, but slightly different for the various temperatures/evolutionary phases
we are sampling. Therefore we applied the \cite{lind} NLTE corrections for  all
our stars, computing these values at the webpage\footnote{
http://inspect-stars.net/} of the INSPECT project. The corrections vary from
about $-$0.09 dex for the warmer MSTO stars to  about $-$0.20 dex for the
coolest RGB stars.  
A large fraction of the stars observed with HIRES and Hydra
(\S\ref{hydra}) are RC and RGB stars near the RC luminosity level;  for these
stars the NLTE corrections are about $-$0.12 dex. Even though the shifts from
LTE to NLTE abundances are not large,  we take them into account to mute
spurious Na abundance trends with \teff\ or \logg\ (see \S\ref{paperend}). We
list both LTE and NLTE abundances of Na in Table~\ref{tab-abund}.

\subsection{Abundance Uncertainties}
Estimates of uncertainties in the abundances began with 
goodness-of-fit trials in matching observed to synthesized spectra.  
For single spectral features, such as the difficult [\ion{O}{1}] transition,
the uncertainties could be as large as 0.1~dex, due to the high noise
level of the HIRES spectra.
These uncertainties decreased as the number of transitions grew (two for
\ion{Na}{1}, up to six for \ion{Ca}{1}, many for C$_2$ and CN).
Likely uncertainties in atmospheric parameters of NGC~6791 stars are
$\pm$150~K in \teff, $\pm$0.2 in \logg, $\pm$0.2~\kmsec\ in \vmicro, and
$\pm$0.1 dex in [Fe/H].
For [\ion{O}{1}], these $\Delta$\teff, $\Delta$\logg, $\Delta$\vmicro,
$\Delta$[Fe/H] excursions lead to $\Delta$~log~$\epsilon$(O) values of 
$\pm$0.03, $\pm$0.12, $\pm$0.00, $\mp$0.03, respectively.
For \ion{Na}{1}, these abundance uncertainties are 
$\pm$0.07, $\mp$0.02, $\mp$0.08, and $\pm$0.02.
The response of C$_2$ and CN to parameter changes is more complex, due
to CO competition in the lower-temperature stars.  
In the coolest RGB stars, increasing \teff\ by 150~K yields smaller
C and N abundances by $\sim$0.10~dex because lessened CO formation 
leaves more free C and N atoms to form C$_2$ and CN.
For the MSTO stars, CO formation is very small, so the same \teff\ increase
creates fewer C$_2$ and CN molecules, raising the C and N abundances
by $\sim$0.10~dex.
Increasing \logg\ by 0.2~dex has no effect on C and N abundances in the
warmer stars and increases them by $\sim$0.05 dex in the cooler ones.
There is little dependence of C$_2$ and CN formation on changes in
\vmicro\ or [Fe/H].
There is a relatively smooth transition between these extreme \teff\ and
\logg\ cases.
The model parameter uncertainties, if viewed as independent variables,
lead to ``external'' abundance uncertainties of about $\pm$0.15 dex.
Thus microturbulent velocity choice is a significant problem in deriving 
reliable Na abundances in NGC~6791 stars.
But in a given region of the CMD all stars will
respond to parameter variations in the same way, and measurement
scatter will be the dominant uncertainty source.

\section{N, NA, AND CA ABUNDANCES FROM HYDRA SPECTRA\label{hydra}} 

To obtain light element abundances for a larger set 
of NGC~6791 stars, we employed the Hydra multifiber spectrograph operating 
at the WIYN telescope on Kitt Peak.
Although we experienced inclement weather on our 2-night observing run,
we obtained enough data to derive  Na and Ca abundances and set boundaries on N for 
21 cluster members (one in common with the HIRES sample).

\subsection{The Hydra Dataset\label{hydraselect}}

To select suitable NGC~6791 targets we began with the 
recent cluster membership survey (\citealt{platais11}) based on proper 
motions, data from which was kindly provided to us by Kyle Cudworth in advance of publication.
We again adopted the star numbering system and photometry of \cite{stetson}.
The proper motions and published radial velocities (RVs) were combined to
identify stars that have high cluster membership probabilities and are bright 
enough ($V$~$<$~16) for the high spectroscopic resolution mode of Hydra.
Photometric $BV$ magnitudes from \citeauthor{stetson} and $K$ magnitudes
from the 2MASS catalog \citep{2mass} are given in Table~\ref{tab-hydra}.
We also list RVs for the targets derived from our spectra (which are used 
to help refine cluster membership probabilities), and we note stars in 
common with previous high-resolution spectroscopic studies.  
Those prior investigations include \citet{carraro06}, \citet{geisler}, 
and \citet[who analyzed the same stars of  \citealt{gra6791}]{car6791}. 
We have no stars in common with \cite{pg98} and \cite{origlia}.
Further discussion of the RVs is given below.

We gathered data for these stars with the upgraded Hydra,  a multifiber
spectrograph operating at the WIYN telescope on Kitt Peak.    We used grating
316@63.4 to obtain spectra near the \ion{Na}{1}  6154, 6160~\AA\ doublet and the
[\ion{O}{1}] 6300~\AA\ line.  The wavelength coverage was 6100-6400~\AA, with a
resolving power  $R \equiv \lambda/\Delta\lambda \simeq 20000$.   We put 26
fibers on target stars and 60 on sky positions.   We observed on UT June 6-7
2009, but only the first night produced  usable data. Even on that night the sky
conditions ranged from mostly clear to very cloudy. This had a direct impact on
the quality of our spectra, with final  signal-to-noise values (see below)
S/N~$\lesssim$ 40.  The total exposure time was 4.5 hours, divided in six
exposures of  2700 seconds each.   We also observed a bright standard star
($\eta$ UMa)  that was later used to cancel the telluric O$_2$ and H$_2$O lines 
from the stellar spectra.

The majority of data reduction tasks were 
accomplished with standard IRAF\footnote{ 
IRAF is distributed by NOAO, which is operated by AURA, under cooperative 
agreement with the NSF.}
routines.  Corrections for bias, zero level, and flat field were applied to the 
individual frames.   We then stacked up all the two-dimensional frames and
extracted the  one-dimensional spectra.   We had to discard one of the targets
because its spectrum signal was  too weak even in the stacked frame.   We
averaged the sky fibers and subtracted the one-dimensional average  sky from the
star spectra.   Wavelength calibration was performed using Th-Ar lamp exposures.
Division by a properly scaled standard star was used to eliminate  the telluric
lines.

We measured RVs using about 40 lines with the  {\sc rvidlines} routine in IRAF. 
Individual heliocentric RVs and associated uncertainties are given in
Table~\ref{tab-hydra}; typical errors are $\simeq$0.5~\kmsec. The identification
of obvious non-members was straightforward, with  three stars clearly not
belonging to the cluster.  One more star (6288) has RV~=~$-55.3$~\kmsec, which
is 6$\sigma$ away  from the average cluster velocity as defined by the other 21
stars  ($<RV>$~= $-$46.3~\kmsec, rms~=~1.5~\kmsec).    This star may be a
non-member, but its spectrum is fully compatible with  the same very high
metallicity of the cluster stars, a value unusual  for a field star. 
Furthermore, it is included in the sample of cluster stars in  \cite{geisler},
who, however, do not publish RV values. Possibly it is a binary, but we have no
way to confirm it.   We decided to keep 6288 among the cluster members.   We had
to exclude one further object (star 5796) from the analysis because  the spectrum
was of too poor quality; this star however is part of the HIRES sample discussed
in \S\ref{keck}.  We were then left with a sample of 21 stars, one of which
was also observed with HIRES (star 7347, see Table~\ref{tab-hydra}).

Table~\ref{tab-hydra} lists S/N estimates for each star.   
There are no extended continuum regions in our line-rich spectra to allow 
easy noise estimation, so we computed S/N in two ways for each star.   
First, we multiplied the count rates in the final spectra by the Hydra 
CCD gain to produce a total electron signal estimate, and used the square 
root of the high points in these counts as one value of S/N. 
Second, we produced a high S/N mean spectrum of ten RC stars\footnote{
Stars contributing to this average were 3712, 4482, 7540, 7912,
8395, 9462, 10695, 11938, 12333, and 12823; see Sect.~\ref{hydraatm} for 
additional use of the RC mean spectrum.}
and subtracted that mean spectrum from those of the individual stars.  In most
cases this resulted in reasonably clean "noise" spectra, from  which we then
computed S/N.  In general the two S/N estimates correlated well.  Each method
has its own  limitations, and so we simply averaged the two values for each
star. These are the numbers reported in Table~\ref{tab-hydra}.

Figure~\ref{NEWcmd} identifies the Hydra targets on the 
CMD of the cluster.  
We concentrated on RGB and RC stars, but put also a few fibers on other 
possibly interesting objects.  
In particular, following the studies by \cite{pg98}, who observed one 
possible horizontal branch (HB) star, and by \cite{liebert}, who demonstrated 
that the extreme blue HB stars are members, we tried to find other HB stars.  
However, none of them has an RV compatible with being part of NGC~6791.  
On the other hand, the candidate Blue Straggler (BSS) star 8481 seems 
to be a true member. 
Note also that since we began our study, the possible HB star studied by 
\cite{pg98} has been convincingly classified as a BSS by \citet{brogaard12}.

\subsection{Atmospheric Parameters and Metallicity for the Hydra
            Sample\label{hydraatm}}

The spectral resolution, $S/N$, and wavelength range values of our Hydra 
spectra are of course less favorable than those of the HIRES spectra. The Hydra
spectra lack the wavelength coverage and  resolution to allow reliable
simultaneous determination of  \teff,  \logg, [M/H], and \vmicro\ for NGC~6791
giants.  Assessment of especially \logg\ cannot be accomplished with our data,
due to the lack of strong first-ion species transitions.  Additionally, there is
a degeneracy between \teff\ and \vmicro\ changes;  variations in these two
parameters cannot easily be disentangled.

Therefore, following our procedures for the HIRES  data (\S\ref{keck}), we
derived temperatures and gravities for the Hydra  stars from their photometry. 
With these two parameters fixed for the stars, we estimated values of  [M/H] and
\vmicro\ from our Hydra data, through minimization of the  difference ($O-C$)
between the observed and synthetic spectra. To accomplish this task, as in
\S\ref{keckcnona} we first created a line list for the entire  6090$-$6400~\AA\
wavelength range of our spectra, beginning with the new CN lines lists of
\cite{bro13b} and  \cite{sne14}.
To these we added atomic lines from the \cite{kur09} line compendium, and 
adjusted their transition probabilities to best match the spectra of Arcturus.

We then applied this linelist to the NGC~6791 program stars.  Because of the
modest $S/N$ and $R$ values of the Hydra spectra, the only  species of the CNO
element group that could be reliably detected was CN.  The [\ion{O}{1}] line was
present but it proved to be  very difficult to analyze; we will comment more on
this problem  in \S~\ref{nacahydra}. All of our Hydra sample are RGB and RC
stars except star 8481, which is a BSS (see Fig.~\ref{NEWcmd}), and which was
not analyzed.  We therefore initially adopted the mean of the CNO abundances
derived  for the three RGB stars (5796, 7347, and 8351) observed with HIRES:
log~$\epsilon$(C)~=~8.73, log~$\epsilon$(N)~=~8.36, and 
log~$\epsilon$(O)~=~8.94 ($\sigma$~$\simeq$~0.05 for each of these  values;
Table~\ref{tab-abund}).

Then we generated grids of synthetic spectra for metallicities [M/H]~=~$-$0.2,
$+$0.1, $+$0.4, $+$0.7, $+$1.0 (where ``M'' essentially  means Fe and other
Fe-peak elements) and microturbulent velocities \vmicro~=~1.00, 1.25, 1.50,
1.75, 2.00 km~s$^{-1}$. After these spectra were smoothed to account for (mainly) the
spectrograph  slit function, they were compared to the observed spectra.  
Arrays of $O-C$ differences were computed, and simple standard  deviations
($\sigma$) were used to assess the quality of fit.  We used the $\sigma$ minima
to determine best values for [M/H] and \vmicro. This method permitted estimation
of both parameters simultaneously.  However, we again emphasize that the derived
[M/H] depends substantially  on \vmicro\ for these high-metallicity stars
because many NGC~6791 lines  on our spectra are saturated.  Not surprisingly,
the $\sigma$ values had broad minima for each star.

In an attempt to more sharply define typical properties for our program  stars,
we also estimated [M/H] and \vmicro\ for the 10-star  RC mean spectrum described
in \S\ref{hydraselect}.  The RC stars have very similar photometric parameters
and hence have  nearly indistinguishable \teff\ and \logg\ values.  We adopted
\teff~=~4545~K and \logg~=2.40 for this NGC~6791 average  RC spectrum and
repeated the synthetic/observed spectrum matches  described above.  The
resulting $\sigma$($O-C$) values converged to a minimum with parameters
[M/H]~=~+0.40~$\pm$~0.15 and \vmicro~=~1.30~$\pm$~0.15 km~s$^{-1}$.  These metallicity and
microturbulence results were  very consistent the HIRES-based values determined
in \S2, and with  previous studies of NGC~6791 stars 
(\citealt{pg98,wor03,carraro06,gra6791,origlia,car6791,boes09,geisler}).

We tried other numerical experiments which possibly could help to define the
metallicities and microturbulent velocities of our stars.  In particular, we
isolated several small spectral regions that contained  mostly \ion{Fe}{1}
lines, and did $O-C$ comparisons with only Fe  abundance variations.  The
results were little different than those from consideration of the  entire
spectral domain and variations in all elements.  Nor did variations in the
synthetic spectrum smoothing function produce  substantial changes in the
overall results.  In the end, for all Hydra stars we chose to adopt the 
\vmicro(\logg) relationship used for the HIRES sample (\S2.2) and a fixed 
metallicity [M/H]~=~+0.4 .

\subsection{Light Elements Abundances from the Hydra Spectra\label{nacahydra}}

We computed synthetic spectra with variations  in N, Na, and Ca abundances,
while keeping C and O abundances fixed and setting and all other abundances at
[X/H]~=~$+$0.4. Synthetic spectrum line lists were constructed in the same
manner as others described previously. As illustrated in Figure~\ref{hydrasyns},
the two \ion{Na}{1} and  four \ion{Ca}{1} lines, and many CN features can easily
be seen; the major problems are line crowding and the modest $S/N$ of even the
best of our Hydra NGC~6791 spectra.

The syntheses of multiple regions rich in CN lines  strongly 
suggest that if C and O have no large star-to-star 
variations, then N is also relatively constant.
We do not list N abundances in Table~\ref{tab-abund} because of their severe
dependence on the other two light elements. If our adopted C and O abundances
given in \S\ref{hydraatm} are essentially correct for all RGB stars, then we
derive a mean abundance $<$log~$\epsilon$(N)$>$~=~8.30 ($\sigma$~=~0.05, 15
stars). This value is very close to the mean N abundance  (8.36) derived for the three
HIRES giants. What is more clear is that we detect no substantial star-to-star
variations in CN strengths beyond those changes induced by variations in \teff\
and \logg\ among NGC~6791 giants. There are no obvious CN-weak or CN-strong
stars in our Hydra sample.

As for the HIRES stars, we derived Na abundances from synthetic spectrum 
computations for the \ion{Na}{1} 6154.2 and 6160.8~\AA\ lines, and Ca abundances
from the \ion{Ca}{1} lines at 6102.7, 6122.2, 6161.3,  6162.2, 6169.0, and
6169.6~\AA. NLTE corrections were applied to the Hydra Na abundances as was done
for the HIRES sample. Simple means of these line abundances were computed, and
those are listed in Table~\ref{tab-abund}. Slightly larger abundances were
derived for the  longer-wavelength \ion{Na}{1} line: 
$<$log~$\epsilon$(6160)~$-$~log~$\epsilon$(6154)$>$ = $+$0.07 
($\sigma$~=~0.04). We attribute this difference to the difficulty in separating
the absorption of the 6160~\AA\ from its \ion{Ca}{1} near neighbors.

%The [\ion{O}{1}] 6300~\AA\ line was 
%substantially compromised by telluric absorption, and its strength was 
%always weak.
Unfortunately our WIYN observing run occurred on non-optimum dates, 
producing a geocentric radial velocity for the cluster that contaminates 
the weak stellar [\ion{O}{1}] line 
with a telluric O$_{\rm 2}$ feature. 
Beyond this problem and the stellar contamination by CN and \ion{Ni}{1}
features coincident with the [\ion{O}{1}] line, at the Hydra effective
resolution there is also substantial blending with the neighboring
\ion{Sc}{2} line at 6300.68~\AA.
For all of these reasons it was not possible to derive a reliable measure
of the O abundance from the Hydra spectra.
However, the [\ion{O}{1}] features are very similar in strength in 
all program stars.   

For the star 7347, in common between the two samples, the
difference in Na and Ca are +0.1, $-0.27$ dex, respectively. However,
comparing the elemental abundances from the two datasets,
for Ca we get 
$<$log $\epsilon$(Hydra)$>$ $-$ $<$log $\epsilon$(HIRES)$>$~= $+$0.01, 
and $+$0.05 if only the three RGB stars from HIRES are included in its mean.
For Na, the difference is
$<$log $\epsilon$(Hydra)$>$ $-$ $<$log $\epsilon$(HIRES)$>$~= $+$0.08,
and $+$0.02 considering only the HIRES RGB stars.
These abundance means are obviously in agreement with one another.

\subsection{Comparison with Other Studies\label{lit}}

NGC~6791 has been the subject of several high-resolution  spectroscopic analyses
\citep{pg98,gra6791,carraro06,origlia,car6791,boes09,geisler,occam}.    All
these papers concur with our determination of a very high metallicity.   The
range in reported metallicities, [Fe/H]=+0.3 to +0.5, is larger than  optimal,
but is consistent with observational/analytical uncertainties.   The
study-to-study variation in individual abundances  
indicates that uncertainties
of about 0.1~dex remain for all abundance  ratios in NGC~6791.

We have stars in common with all of the previous optical studies (see Notes in
Table~\ref{tab-hydra}) except for \cite{pg98}; we  do not have any in common
with the IR study by \cite{origlia} and have not checked the APOGEE spectra,
since only average values are presented in \cite{occam}.  In two cases, we have
analyzed the same HIRES spectra \citep{boes09,geisler}.  Figure~\ref{NEWconf}
shows comparisons between our values for \teff, \logg,  [Fe/H], and Na abundance
and the published ones.

Our temperatures are very close to the ones in  \citet{gra6791} for two RC stars
and hotter by about 100~K than  those in \citet{carraro06} and \citet{geisler};
see panel (a) of  Figure~\ref{NEWconf}. In all these cases, temperatures were
derived from photometry and we consider them to agree well, given the different
colors used.  The larger discrepancy is found for the two MSTO stars in
\cite{boes09},  who derived \teff \ spectroscopically.  The gravities that are
compared in panel (b) of the  figure are very similar for the stars in common
with \cite{gra6791}  and \citet{geisler} and, as already noted in
\S\ref{keckmods},  the differences with asteroseismological results are small. 
For all these stars the agreement is excellent.  Again, there is a larger offset
with \cite{boes09}, who derived gravities  from ionization equilibria and, more
inexplicably, with \citet{carraro06}.

The derived metallicities are also in reasonable  agreement, with differences
below about $\pm0.1$ dex, as illustrated  in panel (c) of Figure~\ref{NEWconf}. 
However, the lower metallicity obtained by \cite{boes09}  for the two MSTO stars
is difficult to understand; we would have expected  an abundance slightly larger
than ours, considering the sensitivity of metallicity to changes in \teff,
\logg, and \vmicro.

Finally, in panel (d) of Figure~\ref{NEWconf} we compare our LTE Na abundances
with those in the literature. We use here the LTE values, since those are
available for all papers. We find that our Na abundances differ by no more than 
about $\pm0.2$~dex from literature values, without trends with  gravity (see
Figure~\ref{NEWconf}) or temperature.

\section{DISCUSSION \label{paperend}} 

Some photometric studies of NGC~6791 have suggested
that this is an unusual OC.
In particular, it hosts blue HB stars (the only case in an open cluster, see
e.g., \citealt{kr,buzzoni}) and its RGB seems broad in color.
Recently, \cite{twarog} raised the idea of a spread in age of 
about 1 Gyr between the inner and outer regions of the cluster, suggested
by the different colors of stars near the MS turn-off.
They were cautious and this was presented only as a possible (if preferred) explanation.  However, the papers by \cite{platais11} and especially 
\cite{brogaard11,brogaard12}, with detailed study of differential 
reddening and the excellent agreement with results based on several 
eclipsing binary systems, theoretical stellar models, and 
asteroseismology, seem to exclude the possibility of an age spread.

A strong indicator of multiple populations in some GCs 
is the existence of a bimodal distribution of CN and CH, beyond what can 
be explained by evolutionary mixing.
The case is strengthened in those GCs with CH/CN bimodality extending
into the domain of unevolved MS stars; see, e.g., \cite{cannon}, 
\cite{har03}, and \cite{pancino}.
A low-resolution spectroscopic study of 19 red horizontal-branch members 
of NGC~6791 was conducted  by \cite{hufnagel}.
They found a likely star-to-star CN band strength spread in their
sample, but ``no bimodality in CN strength is apparent from the spectra of
these stars, unlike the bimodalities among the RHB stars of the two disk
globular clusters 47~Tuc and M71.''
 \cite{carrera}, analysing about 100 SDSS spectra of member stars in
NGC~6791 of different evolutionary phases (main sequence, RGB, and RC), found a
significant spread in the strength of the CN band (only the 3839 \AA, not the
4142 \AA\ band), but only a hint of bimodality. On the other hand, \cite{boberg}
seem to reach a different conclusion, also using SDSS spectra. They do not find evidence to support abundance variations in NGC~6791; however, a full paper with details is not yet availble.

Given the lack of definitive indications from CN, we are 
then left with Na and O abundances as possible indicators of  multiple 
populations, as advocated by \cite{geisler}.

\subsection{Na abundances \label{nanlte}} 

The LTE and NLTE Na distributions as functions of  \teff\ 
are presented in panels (a) and (b) Figure~\ref{NEWna2}. 
The HIRES and Hydra results are distinguished by different symbol colors. 
Histograms of Na abundance are displayed in panels (c) and (d), both 
for the whole sample, and for the Hydra and HIRES subsamples. 
The Hydra stars cover a very limited range in
temperature (they are  essentially concentrated around 4500~K) and exhibit a
moderate spread in  Na abundance: $<$log~$\epsilon$$>$~=~6.72 ($\sigma$~=~0.13,
NLTE). The HIRES stars span a large \teff \ interval, from about 4400~K to
5800~K, and have a slightly lower average Na abundance, with a much smaller 
dispersion: $<$log $\epsilon$$>$~=~6.68 ($\sigma$~=~0.06 dex (NLTE). The HIRES
Na abundances also show a trend of Na abundance with \teff,  which is only
partially removed once NLTE corrections are applied. Interestingly, the mean Na
abundance of the three coolest RGB HIRES stars  ($<$log $\epsilon$$>$~=~6.72,
$\sigma$~=~0.02) agrees with the total  Hydra mean, while the Na abundance of
warmer HIRES stars, at the RGB base and MSTO, is slightly lower: $<$log
$\epsilon$$>$~=~6.67  ($\sigma$~= 0.06).

In our data (but also in those of \citealt{geisler}, see \S\ref{comp}), Na seems
anticorrelated with temperature, i.e., with evolutionary phase: the less evolved
stars have lower Na content.  The effect is $\simeq$0.1~dex in NGC~6791 from
the  MSTO to the RGB. However, an evolutionary effect, with Na brought to the
surface by some  extra-mixing episode, probably can be excluded. The work by
\cite{gra00} showed that O and Na do not change on the RGB,  at least for field
old and metal-poor stars.  Furthermore, recent models by \cite{lagarde},
computed for different  metallicities, both with standard prescriptions and
including thermohaline  convection and rotation-induced mixing, do not prescribe
any change in  Na abundance for masses smaller than about 1.5$M_\odot$, well
above those of our stars.

Is the Na abundance anticorrelation with temperature  an indication that we are
seeing the sign of multiple populations, with  Na enhanced, but only in some of
the stars?  If this is the case, it stands in contrast to trends in GCs, where 
star-to-star variations in Na are seen at any given magnitude  (i.e.,
evolutionary phase) from the MS to the RGB tip.  The only known GC where a small
evolutionary effect (extra-mixing added to self-pollution) may have occurred is
possibly M~13 \citep{jp12}, but  in that cluster an evolutionary signature is
only apparent in the  brightest giants.

There are two possible explanations for this trend. First, the Na lines might be
affected by blends that are not properly accounted for in our analysis. Since
lines are saturated, even weak contaminants may lead to significant errors in
the abundances for such metal-rich stars; such  contaminants are most likely to
be important in cool stars than in hot ones. 
In addition to the  blending with the Ca I lines mentioned 
in \S\ref{keckcnona} and \ref{hydraselect}, 
the Na~I lines are also contaminated by CN lines, as illustrated 
by Figure~\ref{hydrasyns}, where CN is responsible for the residual 
absorption at the wavelength of the Na lines in synthetic spectra
computed without Na and Ca contributions.
While these contaminants are
considered in our analysis, there might be others  that we neglected. Second,
there might be problems with the 1-d model atmospheres in reproducing  the Na I
lines, especially for the coolest giants. The 3-d models \citep[e.g.,][]{cobold}
predict steeper temperature gradients than 1-d model atmospheres, the difference
being larger for giants than for dwarfs, and this could also possibly explain
the about 0.1 dex difference we find in Na abundance between the MSTO and
RGB/RC.

\subsection{O and Na \label{nao2}} 

The most important result of our analysis is that we find no abundance 
dispersion exceeding the associated errors for Na and O. We also do not see any
indication of the presence of a Na-O anticorrelation.  In Figure~\ref{NEWnao} we
show a plot of the O and Na abundances derived  in this work for NGC~6791, and
for comparison include also the Na and O abundances derived in our FLAMES survey
of GCs (\citealt{carretta09}).  At variance from the GC data, Na and O
abundances seem  even to show a slight positive correlation in NGC~6791 (however
the  Pearson coefficient is not significant for this sample).  The rms scatters
of both O and Na are compatible with their associated  errors, especially
considering the Hydra and HIRES samples separately.

The abundance of Na have been discussed in \S\ref{nanlte}.  The mean O abundance
in NGC~6791 ([O/Fe]~=~$-$0.2), lies  between the values found by \cite{origlia}
and \cite{car6791}, and is  definitely lower than the values for both
sub-populations (Na-rich and  Na-poor) in \cite{geisler}.  In absence of any
self-pollution like that seen in GCs, a sub-solar ratio  is expected for the
high metallicity of this OC, see e.g., \cite{bensby}.

However, should have we expected a large effect even under the  self-pollution
hypothesis?  There are observational indications that O variations are smaller
for  higher metallicity GCs.  Consider in particular the bulge GC NGC~6553.
\cite{melendez} derived an overall metallicity [Fe/H]~=~$-0.20$, and found
[O/Fe]~=~$+$0.20 ($\sigma$~=~0.065) for five RGB stars observed  in the H band
in this cluster. \cite{cohen99} studied 5 red HB stars in NGC~6553, deriving a
similar  metallicity and higher O abundance ([O/Fe]~=~$+$0.50), but with a
scatter  small enough ($\sigma$~=~0.13) and comparable to the star-to-star
variations found for the other elements which have relatively  constant [X/Fe]
abundance ratios.

Assume for the moment that the present O and Na abundances  reflect the
nucleosynthesis contributions of multiple generations. If the original polluters
(i.e., the stars that lost the material  that produced the second generation of
stars) are AGB stars, the lack of significant star-to-star O abundance scatter
in NGC~6553 and NGC~6791 could  be due to the efficiency of the hot bottom
burning (HBB) mechanism.  At high metallicity the envelope temperatures are
lower, hence the ON  cycle is not activated (D'Antona, private communication). 
Indeed, the AGB models by \cite{vda09} have O yields that do not show any
depletion at the high-metallicity end.  Of course these models still suffer from
many uncertainties, and are  furthermore computed for metallicities much lower
than the one of  NGC~6791 (their higher metallicity model is for Z~=~0.004, or 
[Fe/H]~$\simeq$~$-$0.7), but the trend of the yields with metallicity is clear
(see their Table~2).

\subsection{Comparison with \cite{geisler}'s results \label{comp}}

In the \cite{geisler} study of NGC~6791, a significant 
trend of Na abundance with \teff\ was noted, but no explanation was provided.
They instead concentrated on the bimodality seen in their abundances. 
In Figure~\ref{NEWna2gei} we contrast our Na abundances with their results.
In panel (b) of this figure one can see two groups of Na abundances
separated by $\sim$0.5 dex in their results that is obviously not apparent 
in our data from panel (a).
But trends of Na with \teff\ are visible in both of their [Na/Fe] subsamples.
 In Figure~\ref{newoxy} we show similar plots, but for oxygen. This is less
significant, though, since O does not vary a lot in either sample.

We notice that the large differences obtained between our analyses and those of
\cite{geisler} cannot be attributed to the atmospheric parameters, which are
similar (\S\ref{lit}).  We note that the microturbulent velocity they adopt for
the stars in the lower RGB are  unusually low, but the impact of \vmicro\ on Na
or O abundance is too small to significantly affect  the [Na/Fe] trends. A more
promising candidate is the different way we accounted for nearby  contaminating
lines in the line syntheses, in particular the CN and Ca lines that are quite
strongly temperature sensitive. Dr. Sandro Villanova kindly provided us with
the  list of lines used in their spectral synthesis of the region including the
Na lines. They used different values from those adopted by us; when used in the
synthesis their values produced  much weaker contamination for cool stars, while
the effect was much smaller for the hot ones.

Finally, while there seems to be a sort of Na-O anticorrelation in 
\citet[][see their Figure~4]{geisler}, it is different from what found in GCs.
First, it seems only present (if at all) for the Na-rich stars. 
Second, the Na-poor and Na-rich subsamples seem to have very similar 
O content. 
Third, there seems to be a correlation, such that Na-rich 
stars are also O-rich, at variance with the case of GCs.
If the effect is real, this may be the manifestation of the different 
O production in high-metallicity stars (\S\ref{nao2}).

\section{Summary \label{summary}}

We studied possible light abundance anomalies and spreads in NGC~6791, 
analyzing a sample of 36 MSTO, RGB, and RC stars. 
About one half of the stars were observed with WIYN/Hydra and one 
half retrieved from the Keck HIRES archive; this is the largest sample 
examined up  to now in this cluster.

We concentrated on determining abundances of N, O, and 
Na using spectrum synthesis and EW measurements.
We did not find any significant dispersion in O and  Na, 
nor an anticorrelation between them. 
We found, however, a small dependence of Na abundance on temperature 
(i.e., evolutionary status), only partially removed when NLTE corrections 
are applied.

Based on the available information, both photometric (from literature 
data) and spectroscopic (mainly the present work), NGC~6791  can still 
be considered a bona fide open cluster. 
In other words, it is a very homogeneous 
aggregate, formed with a different mechanism with respect to the
multi-population GCs.
However, further observations and maybe a different analysis method 
(e.g., using 3-d model atmospheres) could help in definitely pinning 
down the cluster's elusive properties.

\acknowledgments
We wish to  thank Kyle Cudworth for his valuable 
information on membership based on proper motions and Andrea Miglio 
for providing information on the Kepler targets in advance of publication
We thank Sandro Villanova for having kindly provided us the line list they used
in the synthesis of the Na line region and one of their spectra in this same
region.
The help of the WIYN telescope personnel is acknowledged as are fruitful 
discussions with Franca D'Antona. 
This research has made use of the Keck Observatory Archive (KOA), which 
is operated by the W. M. Keck Observatory and the Exoplanet Science 
Institute (NExScI), under contract with the U.S. National Aeronautics and 
Space Administration (NASA).
We acknowledge the use of Keck data obtained by the following PIs: 
Boesgaard, Bresolin, and Marcy.
This research makes use of the SIMBAD database, operated at CDS, Strasbourg,
France, of NASA's Astrophysics Data System, and of the WEBDA open cluster
database.  
This publication makes use of data products from the Two Micron All 
Sky Survey, which is a joint project of the University of Massachusetts 
and the Infrared Processing and Analysis Center/California Institute of 
Technology, funded by NASA and the U.S. National Science Foundation (NSF).

This work was supported by PRIN INAF 2011 ``Multiple 
populations in globular clusters: their role in the Galaxy assembly" 
(PI E. Carretta), by PRIN MIUR 2010-2011 ``The chemical and Dynamical 
Evolution of the Milky Way and Local Group Galaxies (PI F. Matteucci), 
and by the U.S. NSF under grants AST-0908978 and AST-1211585.
The paper was completed while C.S. was on a University of Texas 
Faculty Research Assignment, in residence at the Department of Astronomy 
and Space Sciences of Ege University.  
Financial support from the University of Texas and The Scientific and 
Technological Research Council of Turkey (TBITAK, project No. 112T929) 
are greatly appreciated. The research at the University of York has been 
supported by funds from the Leverhulme Trust (UK).  
Some funding was also provided by the NASA laboratory astrophysics program.

{\it Facilities:} \facility{WIYN (Hydra)}, \facility{Keck (HIRES)}.

\clearpage
\begin{figure}
\centering
\includegraphics[scale=0.4]{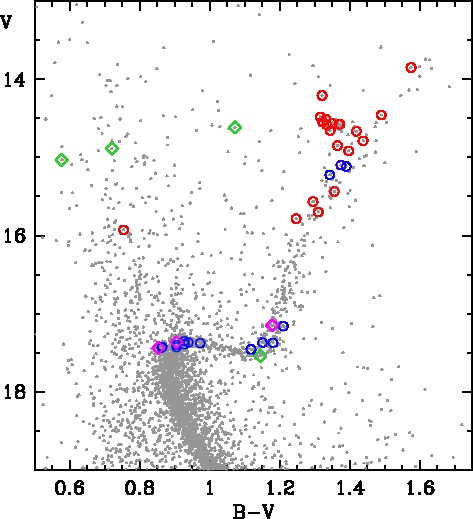} 
\caption{\footnotesize
The $V,B-V$ CMD for NGC~6791  \citep{stetson}, with our targets indicated. 
Large circles indicate cluster members (red for stars observed with Hydra, 
and blue for those observed with HIRES) and diamonds indicate non members 
(green for those excluded from RV mismatches, and magenta from metallicity
discordance).}
\label{NEWcmd}
\end{figure}

\begin{figure}
\centering
\includegraphics[scale=0.4]{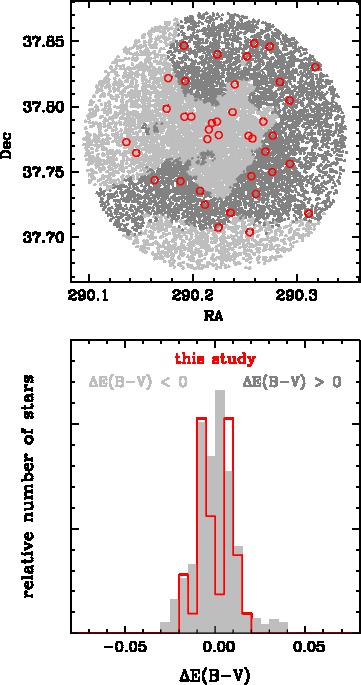} 
\caption{\footnotesize
Upper panel: Map of differential reddening 
\citep{brogaard11} with our targets indicated. 
Darker and lighter gray indicate negative and positive differential 
reddening ($\Delta$E(B$-$V)~$<$~0 and $>$~0), respectively. 
Our target stars from both HIRES and Hydra samples are indicated
with red circles.
Lower panel: Histograms of the differential reddening values for the 
entire sample (gray, filled) and for our targets (open, red).
For ease of comparison, the individual histograms are individually 
normalized to yield approximately the same vertical extents.}
\label{NEWdiffred}
\end{figure}

\begin{figure}
\centering
\includegraphics[scale=0.4]{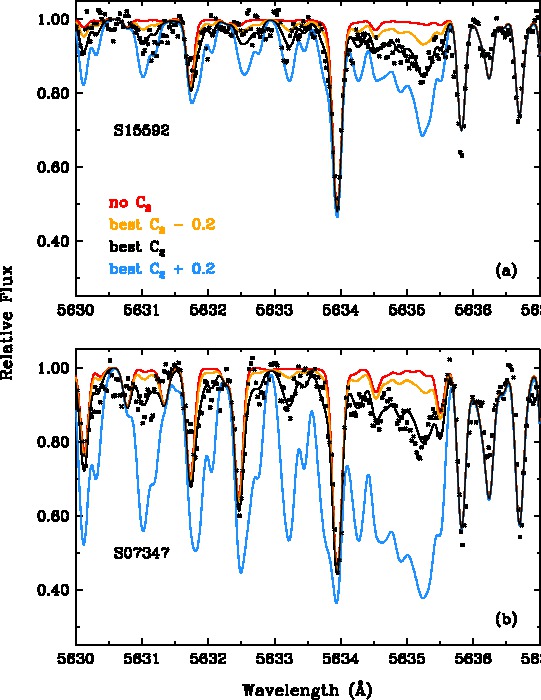}  
\caption{\footnotesize
Observed HIRES and synthetic C$_2$ Swan system 0$-$1 bandhead for two
NGC~6791 stars.
In panel (a) the warm subgiant star 15592 is displayed, and in panel 
(b) the RGB star 7347 is shown.  
The points represent the observed spectra, and the abundances of the four
synthetic spectra are identified in the figure legend.
The ``best'' C$_2$ means that C abundance which best matches the observed
C$_2$ bandhead.}
\label{c2synsc}
\end{figure}

\begin{figure}
\centering
\includegraphics[scale=0.4]{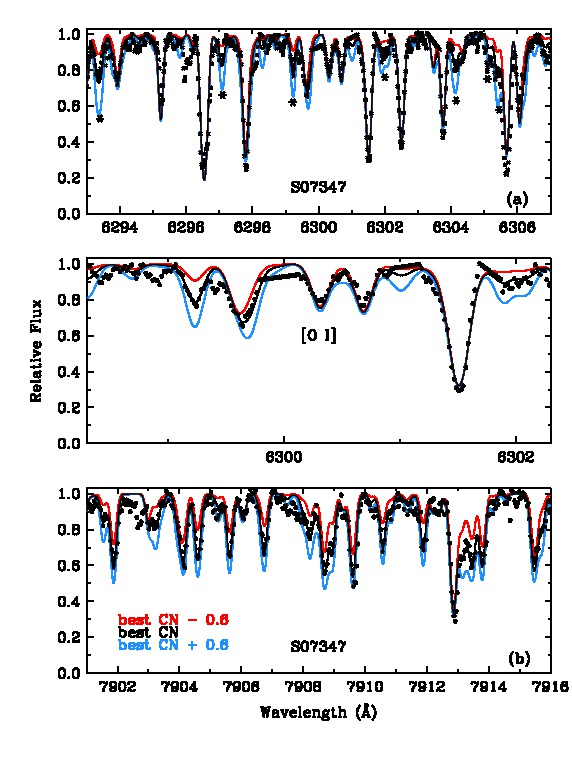}   
\caption{\footnotesize
Observed HIRES and synthetic spectra of NGC~6791 RGB star 7347 in
small spectral regions surrounding the [\ion{O}{1}] 6300.3~\AA\ line 
 (see upper panel and an enlargement in the middle panel) and 
near 7910~\AA. 
The N abundance has been varied to show the presence of CN lines
in the spectrum.
Star 7347 is illustrative of all program stars with \teff~$<$~5000~K.
In the figure, points represent the observed spectra, and the abundances 
of the three synthetic spectra are identified in the figure legend.
Six-pointed stars mark the major CN features.
The ``best'' CN means that N abundance which produces syntheses that
best matches the observed CN lines.}
\label{cnsyns}
\end{figure}

\begin{figure}                                                
\centering                                                    
\includegraphics[scale=0.4]{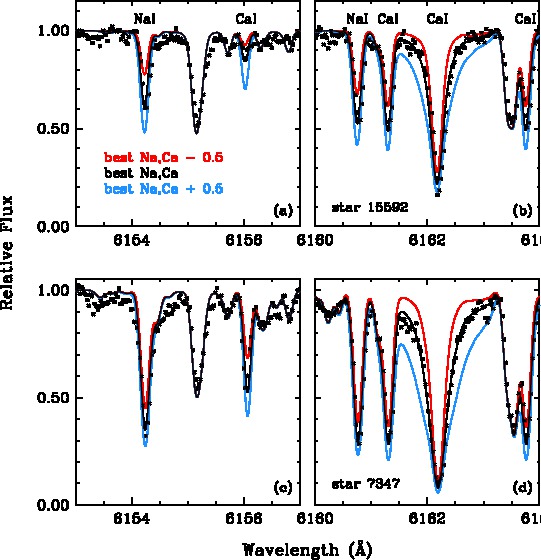}   
\caption{\footnotesize
Observed HIRES and synthetic spectra of two NGC~6791 stars in the
spectral regions surrounding \ion{Na}{1} 6154 and 6160~\AA\ and 
nearby \ion{Ca}{1} lines.
The observed spectra are shown as black points.
As in Figure~\ref{c2synsc}, the warm subgiant star 15592 is shown in 
top panels (a) and (b), and the RGB star 7347 is shown in bottom
panels (c) and (d).
The assumed Na and Ca abundances in the syntheses differ by 0.5 dex 
(as written in the figure legend) in order
to clearly show the abundance effects on the displayed transitions.
In the figure, points represent the observed spectra.}
\label{nacasyns}                                                
\end{figure}

\begin{figure}
\centering 
\includegraphics[scale=0.4]{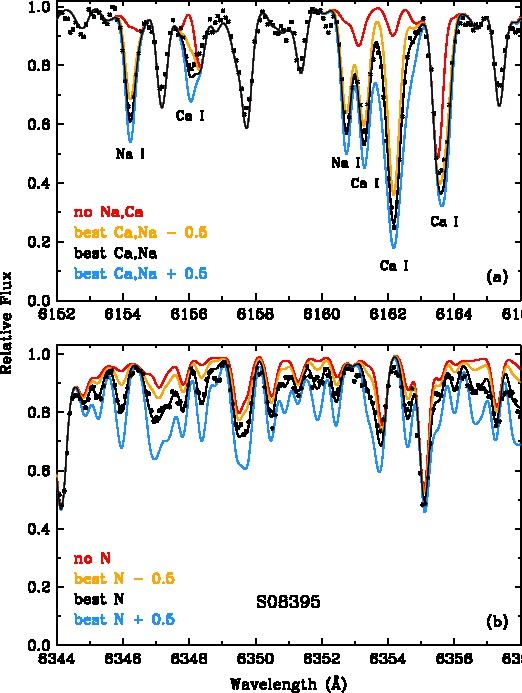}  
\caption{\footnotesize
Observed Hydra and synthetic spectra of NGC~6791 RGB star 8395 in
two small spectral intervals.
Panel (a) shows the region containing the two \ion{Na}{1} lines and
four of the six \ion{Ca}{1} lines that were used to assess Ca abundances
with the Hydra spectra.
Panel (b) shows one of the several regions that have substantial
CN contribution to the total absorption.
As in previous figures, the observed points are indicated with black points,
and the synthetic spectra are drawn with color-coded lines that have
the abundances that are explained in the figure legends.}
\label{hydrasyns}
\end{figure}

\begin{figure}
\centering
\includegraphics[scale=0.4]{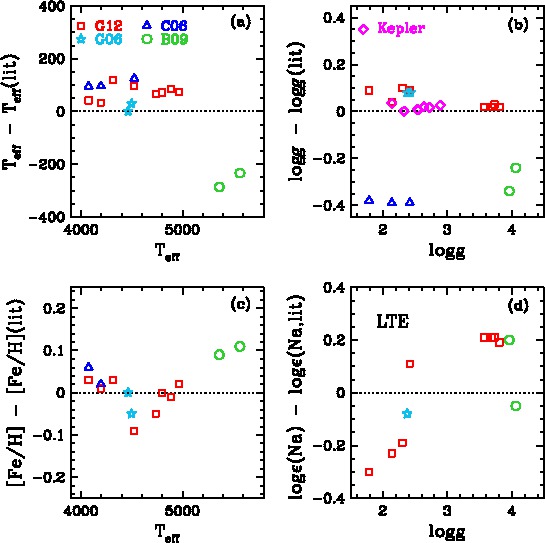}   
\caption{\footnotesize
Comparison between our results and literature values for 
temperature, gravity, metallicity, and Na abundance (LTE case). 
As identified in the legend of panel (a), red open squares are used 
for the results of \citet[][G12]{geisler}, 
blue open triangles for \citet[][C06]{carraro06},
cyan stars (G06) for \cite{gra6791} and \cite{car6791}, 
and green circles for \citet[][B09]{boes09}. 
In panel (b) magenta diamonds are added for \citet[][Kepler]{stello}.
We plot log~$\epsilon$(Na) and not [Na/Fe], in order to isolate the 
effect of Na variation from differences in Fe.}
\label{NEWconf}
\end{figure}

\begin{figure}
\centering
\includegraphics[scale=0.4]{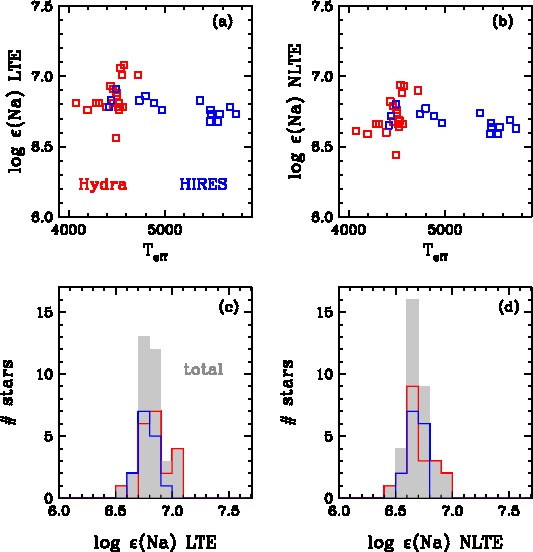}   
\caption{\footnotesize
Na abundances from our HIRES and Hydra spectra.
In all panels red color denotes results from Hydra and blue color
those from HIRES.
The upper panels show Na abundances as a function of temperature for
the LTE (panel a) and NLTE (panel b) cases. 
Adoption of NLTE corrections from \cite{lind} eliminates only part of 
the small trend of Na abundance with temperature. 
The lower panels show the histograms of the Na abundances for the 
total sample (gray, filled) in addition to the individual histograms
of Hydra and HIRES abundances (panels c and d for LTE and NLTE results, 
respectively).}
\label{NEWna2}
\end{figure}

\begin{figure}
\centering
\includegraphics[scale=0.4]{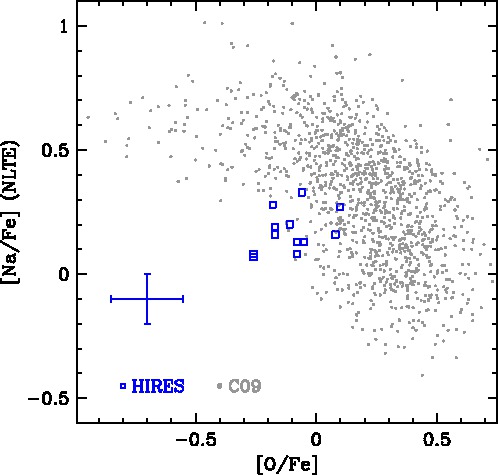}   
\caption{\footnotesize
The Na (NLTE) and O values for our sample in NGC~6791. 
These are shown as blue squares, and are only from the HIRES sample.
 Typical error bars are shown on the lower left of the panel.
For reference, we show as gray points the Na and O values from our FLAMES 
survey of GCs \cite[][G09]{carretta09}.}
\label{NEWnao}
\end{figure}

\begin{figure}
\centering
\includegraphics[scale=0.4]{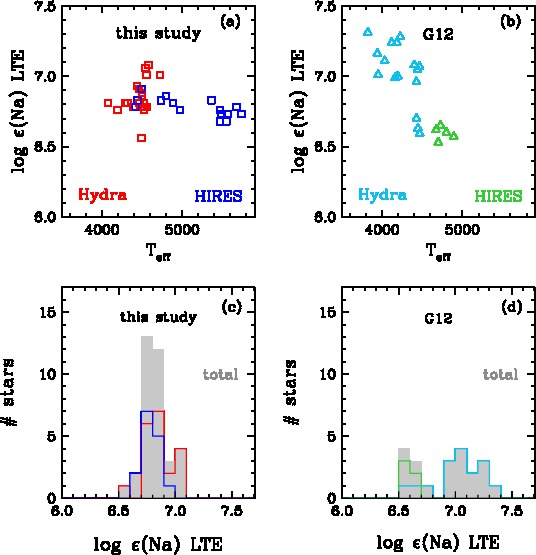}   
\caption{\footnotesize
Comparison of our results for Na with those of \cite{geisler}. 
The left-hand panels (a) and (c) are repeats of those panels in 
Figure~\ref{NEWna2}, while right-hand panels are results from
\citeauthor{geisler} (2102, G12).}
\label{NEWna2gei}
\end{figure}

\begin{figure}
\centering
\includegraphics[scale=0.4]{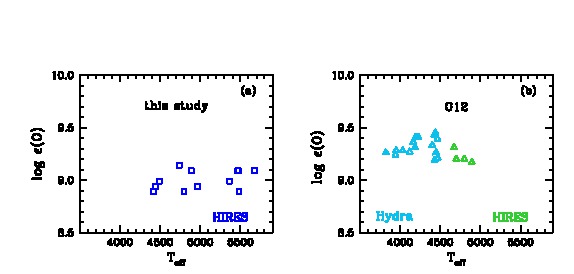}   
\caption{\footnotesize 
Comparison of our results for O with those of \cite{geisler}; symbols are as
in previous figure.}
\label{newoxy}
\end{figure}

\clearpage

\begin{table*}
\centering 
\scriptsize
\caption{The HIRES Sample} 
\setlength{\tabcolsep}{0.6mm}
\begin{tabular}{rccccccccccl}
\tableline\tableline
ID      & RA         & Dec        &  B    &  V    &  K    & S/N  &T$_{\rm eff}$ &$\log$ g &$v_t$     &Prob  &Remarks\\
        &(h:m:s)     &(d:p:s)     &       &       &2MASS  &      &(K)           &     &(km~s$^{-1}$) &(p.m.) &     \\
\tableline
   5744 &19:20:46.08 &+37:47:32.3 &18.350 &17.377 &15.477 & 45  &5364 &3.96 &0.94 & 90  & HIRESa, publ.\\
   7649 &19:20:51.71 &+37:46:57.0 &18.333 &17.428 &15.475 & 16  &5566 &4.06 &0.91 & 78  & HIRESa, publ.\\
  11220 &19:21:02.19 &+37:50:54.0 &18.278 &17.350 &15.190 & 33  &5468 &3.99 &0.93 &\dots& HIRESa\\
  12181 &19:21:05.72 &+37:50:45.6 &18.305 &17.400 &15.763 & 32  &5551 &4.05 &0.92 &\dots& HIRESa\\
  13334 &19:21:10.37 &+37:48:17.6 &18.306 &17.367 &15.497 & 31  &5474 &4.00 &0.93 &\dots& HIRESa\\
  13352 &19:21:10.40 &+37:45:22.6 &18.298 &17.435 &15.128 & 36  &5738 &4.13 &0.89 &\dots& HIRESa\\ 
  14416 &19:21:16.28 &+37:49:49.5 &18.319 &17.391 &15.118 & 40  &5482 &4.01 &0.93 &\dots& HIRESa\\
  15592 &19:21:25.71 &+37:46:53.3 &18.29  &17.425 &15.397 & 42  &5675 &4.10 &0.90 &\dots& HIRESa\\
   9609 &19:20:57.20 &+37:47:45.0 &18.368 &17.158 &14.357 & 25  &4738 &3.57 &1.07 & 53  & HIRESb, T32\\
  11014 &19:21:01.49 &+37:44:48.6 &18.575 &17.457 &14.904 & 29  &4968 &3.81 &0.99 &\dots& HIRESb, T33\\
  11092 &19:21:01.76 &+37:46:32.3 &18.553 &17.372 &14.623 & 32  &4799 &3.69 &1.03 & 88  & HIRESb, T34\\
  12382 &19:21:06.45 &+37:46:40.1 &18.520 &17.369 &14.786 & 31  &4887 &3.73 &1.02 &\dots& HIRESb, T35\\
   5796 &19:20:46.25 &+37:49:10.8 &16.566 &15.223 &12.144 & 30  &4493 &2.64 &1.37 & 99  & HIRESc, Hydra, Kepler\\
   7347 &19:20:50.83 &+37:43:29.5 &16.474 &15.099 &11.935 & 34  &4444 &2.55 &1.40 & 99  & HIRESc, Hydra, Kepler\\
   8351 &19:20:53.59 &+37:47:19.0 &16.507 &15.117 &11.920 & 34  &4418 &2.54 &1.40 & 96  & HIRESc, Kepler\\
   4591 &19:20:42.34 &+37:49:18.7 &18.292 &17.439 &15.081 &\dots&\dots&\dots&\dots&\dots& HIRESa, NM\\
   2130 &19:20:31.47 &+37:54:26.3 &18.27  &17.365 &15.112 &\dots&\dots&\dots&\dots&\dots& HIRESa, NM?\\
   8506 &19:20:53.99 &+37:46:41.9 &18.329 &17.150 &14.335 &\dots&\dots&\dots&\dots&\dots& HIRESb, NM\\
   6509 &19:20:48.37 &+37:46:30.2 &18.687 &17.542 &14.440 &\dots&\dots&\dots&\dots&\dots& HIRESb, NM\\
\tableline
\end{tabular}
\tablecomments{ID, B, V from \cite{stetson}; coordinates and Prob (p.m.) from Cudworth (priv.comm.).\\ 
Kepler indicates that the star has been observed by the $Kepler$ satellite and is considered an astroseismologic
 member \citep{stello}.\\
T32, T33, T34, T35 are identifications in \cite{geisler}; NM=not a member \\
PIs of Keck programs: HIRESa : A.M. Boesgaard (U. Hawaii), two stars published in \cite{boes09}; HIRESb: F. Bresolin (U. Hawaii),
published in \cite{geisler}; HIRESc: G.W. Marcy (U. California Berkeley)} 
\label{tab-hires}
\end{table*}

\begin{table*}
\centering \scriptsize
\caption{The Hydra Sample} 
\setlength{\tabcolsep}{0.6mm}
\begin{tabular}{rccccccccccccl}
\tableline\tableline
ID      & RA         & Dec        &  B    &  V    &  K    & S/N  & RV    & err &T$_{\rm eff}$ &$\log$ g &$v_t$       &Prob  &Remarks\\
        &(h:m:s)     &(d:p:s)     &       &       &2MASS  &      &(km~s$^{-1}$) &(km~s$^{-1}$) &(K) &         &(km~s$^{-1}$) &(p.m.) &     \\
\tableline
   2309 &19:20:32.66 &+37:46:22.4 &16.226 &14.788 &11.424 & 26 &-47.81 &0.50 &4283 &2.33 & 1.47 & 33 & Kepler	    \\
   3712 &19:20:39.23 &+37:44:37.5 &16.213 &14.848 &11.827 & 15 &-44.38 &0.62 &4502 &2.51 & 1.41 & 98 &      \\
   4482 &19:20:41.97 &+37:47:54.4 &15.871 &14.548 &11.517 & 10 &-43.85 &0.78 &4530 &2.31 & 1.44 & 98 & rebinned,Kepler      \\
   5454 &19:20:45.18 &+37:44:34.2 &16.312 &14.915 &11.704 & 25 &-47.59 &0.40 &4393 &2.47 & 1.43 & 98 &      \\
   5706 &19:20:45.98 &+37:50:48.3 &17.033 &15.786 &13.031 & 11 &-44.64 &0.83 &4723 &3.02 & 1.25 & 85 & 	 \\
   6288 &19:20:47.66 &+37:47:32.5 &16.084 &14.665 &11.267 & 12 &-55.33 &0.40 &4315 &2.31 & 1.48 & 29 & bin/NM?, G12, Kepler   \\
   6940 &19:20:49.67 &+37:44:08.0 &15.923 &14.588 &11.500 & 25 &-44.61 &0.37 &4525 &2.42 & 1.44 &  0 & C06, G12     \\
   7347 &19:20:50.83 &+37:43:29.5 &16.474 &15.099 &11.935 & 13 &-44.70 &0.43 &4435 &2.57 & 1.39 & 99 & rebinned, Kepler, HIRES  \\
   7540 &19:20:51.38 &+37:46:30.7 &15.527 &14.206 &11.246 & 41 &-47.05 &0.42 &4578 &2.30 & 1.48 & 51 &      \\
   7912 &19:20:52.44 &+37:47:15.4 &17.009 &15.699 &12.700 & 11 &-49.59 &0.73 &4554 &2.89 & 1.29 & 97 & rebinned, Kepler  \\
   8395 &19:20:53.69 &+37:50:24.0 &15.943 &14.574 &11.549 & 36 &-47.85 &0.52 &4494 &2.40 & 1.45 & 89 & Kepler	    \\
   8481 &19:20:53.92 &+37:42:26.3 &16.681 &15.927 &14.251 & 10 &-47.18 &0.69 &6178 &3.77 &\dots & 97 & BSS \\
   9462 &19:20:56.72 &+37:43:07.8 &16.000 &14.656 &11.586 & 26 &-45.86 &0.51 &4498 &2.43 & 1.44 & 99 & G06 (3019), Kepler\\
   9786 &19:20:57.71 &+37:49:01.5 &16.791 &15.435 &12.336 & 24 &-46.46 &0.40 &4463 &2.72 & 1.34 & 85 & Kepler	    \\
  10695 &19:21:00.53 &+37:50:19.1 &15.843 &14.511 &11.493 & 29 &-46.19 &0.38 &4524 &2.30 & 1.45 & 88 & Kepler	 \\
  10806 &19:21:00.87 &+37:46:39.9 &15.948 &14.576 &11.468 & 34 &-47.60 &0.41 &4464 &2.38 & 1.45 & 97 & G06 (2014), Kepler\\
  10898 &19:21:01.13 &+37:42:13.8 &15.949 &14.459 &10.938 & 30 &-47.20 &0.42 &4196 &2.13 & 1.53 & 68 & C06, G12, Kepler     \\
  11814 &19:21:04.27 &+37:47:18.9 &15.424 &13.849 &10.102 & 34 &-46.57 &0.43 &4075 &1.79 & 1.64 & 98 & C06, G12, Kepler     \\
  11938 &19:21:04.74 &+37:45:56.4 &15.914 &14.564 &11.527 & 31 &-46.27 &0.39 &4511 &2.40 & 1.45 & 96 & Kepler	 \\
  12333 &19:21:06.31 &+37:44:59.9 &15.799 &14.483 &11.499 & 20 &-44.28 &0.46 &4563 &2.41 & 1.45 & 99 & Kepler	    \\
  12823 &19:21:08.07 &+37:49:08.6 &16.862 &15.567 &12.554 & 19 &-46.56 &0.69 &4539 &2.82 & 1.31 & 78 &      \\
\multicolumn{14}{c}{Not analyzed / Non members} \\
   5796 &19:20:46.25 &+37:49:10.8 &16.566 &15.223 &12.144 &\dots&-45.34 &0.91 &\dots&\dots&\dots& 99  & HIRES, Very low S/N, Kepler\\
   2746 &19:20:34.97 &+37:45:52.5 &15.609 &15.031 &13.833 &\dots&-15.71 &0.89 &\dots&\dots&\dots& 89  & NM	       \\     
  11335 &19:21:02.56 &+37:43:59.8 &15.606 &14.886 &13.195 &\dots& 15.66 &0.34 &\dots&\dots&\dots&  2  & NM	       \\     
  14153 &19:21:14.75 &+37:43:05.6 &15.687 &14.615 &12.047 &\dots&-67.20 &0.37 &\dots&\dots&\dots&\dots& NM      \\  
\tableline
\end{tabular}

\tablecomments{ID, B, V from \cite{stetson}, except B for stars 2130 and 15592,
average between \cite{mjp} and \cite{kr}; coordinates and Prob (p.m.) from Cudworth (priv.comm.). 
C06: \cite[who use Stetson's IDs]{carraro06}.
G06: \cite[names indicated in Remarks]{gra6791}; G12: \cite{geisler}; Kepler indicates that the star has been observed
by the $Kepler$ satellite and is considered an astroseismologic member \citep{stello}.\\
NM=not a member for RV} 
\label{tab-hydra}
\end{table*}

\clearpage

\begin{table*}
\centering 
\scriptsize
\caption{Abundances for the HIRES and Hydra stars}
\vspace*{3mm}
\setlength{\tabcolsep}{1.5mm}
\begin{tabular}{rcccccccc cccccccc}
\hline
Star  & Fe  &C    &N 	 &O    &Ca  &Ni   &Na	&Na   &[Fe/H] &[C/Fe] &[N/Fe] &[O/Fe] &[Ca/Fe] &[Ni/Fe] &[Na/Fe] &[Na/Fe]\\
      &     &	  &	 &     &    &	  &LTE  &NLTE &       &	      &       &       &        &        &LTE     &NLTE   \\
\hline
\multicolumn{17}{c}{Keck stars}\\      
 5744 &7.91 &8.88 &\dots&8.99 &6.66 &6.80 &6.83 &6.74 & 0.37& -0.01 & \dots & -0.17 &  0.04 &  0.15 & 0.25 &  0.16 \\	 
 7649 &7.96 &\dots&\dots&\dots&6.89 &6.90 &6.73 &6.64 & 0.42& \dots & \dots & \dots &  0.22 &  0.20 & 0.10 &  0.01 \\ 
11220 &7.76 &8.83 &\dots&9.09 &6.48 &6.74 &6.68 &6.59 & 0.22&  0.11 & \dots &  0.08 &  0.01 &  0.24 & 0.25 &  0.16 \\	  
12181 &7.93 &8.88 &\dots&\dots&6.61 &6.77 &6.68 &6.59 & 0.39& -0.02 & \dots & \dots & -0.03 &  0.10 & 0.08 & -0.01 \\	 
13334 &7.92 &8.83 &\dots&9.09 &6.60 &6.65 &6.76 &6.67 & 0.38& -0.05 & \dots & -0.08 & -0.03 & -0.01 & 0.17 &  0.08 \\	 
13352 &7.91 &8.95 &\dots&\dots&6.75 &6.72 &6.73 &6.63 & 0.37&  0.08 & \dots & \dots &  0.13 &  0.07 & 0.15 &  0.05 \\	 
14416 &7.90 &8.90 &\dots&8.89 &6.66 &6.65 &6.73 &6.64 & 0.36&  0.04 & \dots & -0.26 &  0.05 &  0.01 & 0.16 &  0.07 \\	 
15592 &7.89 &8.90 &\dots&9.09 &6.67 &6.69 &6.78 &6.69 & 0.35&  0.05 & \dots & -0.05 &  0.07 &  0.06 & 0.22 &  0.13 \\
 9609 &7.79 &8.88 &8.32 &9.14 &6.48 &6.74 &6.83 &6.73 & 0.25&  0.13 &  0.13 &  0.10 & -0.02 &  0.21 & 0.37 &  0.27 \\	 
11014 &7.80 &8.83 &8.36 &8.94 &6.54 &6.80 &6.76 &6.67 & 0.26&  0.07 &  0.18 & -0.11 &  0.03 &  0.26 & 0.29 &  0.20 \\	 
11092 &7.82 &8.78 &8.3  &8.89 &6.50 &6.74 &6.86 &6.77 & 0.28&  0.00 &  0.10 & -0.18 & -0.03 &  0.18 & 0.37 &  0.28 \\	 
12382 &7.92 &8.91 &8.29 &9.09 &6.43 &6.71 &6.81 &6.72 & 0.38&  0.03 & -0.01 & -0.08 & -0.20 &  0.05 & 0.22 &  0.13 \\	 
 5796 &7.80 &8.73 &8.46 &8.99 &6.59 &6.73 &6.91 &6.80 & 0.26& -0.03 &  0.28 & -0.06 &  0.08 &  0.19 & 0.44 &  0.33 \\
 7347 &7.86 &8.75 &8.41 &8.94 &6.51 &6.61 &6.83 &6.72 & 0.32& -0.07 &  0.17 & -0.17 & -0.06 &  0.01 & 0.30 &  0.19 \\	 
 8351 &7.90 &8.70 &8.36 &8.89 &6.54 &6.74 &6.78 &6.65 & 0.36& -0.16 &  0.08 & -0.26 & -0.07 &  0.10 & 0.21 &  0.08 \\		
\multicolumn{17}{c}{Hydra stars}\\	   	          	     
 2309 &\dots&\dots&\dots&\dots&6.59 &\dots&6.81 &6.66 &+0.4 &\dots&\dots&\dots&-0.14&\dots& 0.20& 0.05 \\
 3712 &\dots&\dots&\dots&\dots&6.62 &\dots&6.86 &6.74 &+0.4 &\dots&\dots&\dots&-0.06&\dots& 0.25& 0.13 \\
 4482 &\dots&\dots&\dots&\dots&6.75 &\dots&6.78 &6.68 &+0.4 &\dots&\dots&\dots& 0.11&\dots& 0.17& 0.07 \\
 5454 &\dots&\dots&\dots&\dots&6.59 &\dots&6.78 &6.60 &+0.4 &\dots&\dots&\dots&-0.10&\dots& 0.17&-0.01 \\
 5706 &\dots&\dots&\dots&\dots&6.62 &\dots&7.01 &6.90 &+0.4 &\dots&\dots&\dots&-0.04&\dots& 0.40& 0.29 \\
 6288 &\dots&\dots&\dots&\dots&6.49 &\dots&6.81 &6.66 &+0.4 &\dots&\dots&\dots&-0.06&\dots& 0.20& 0.05 \\
 6940 &\dots&\dots&\dots&\dots&6.62 &\dots&6.81 &6.69 &+0.4 &\dots&\dots&\dots&-0.08&\dots& 0.20& 0.08 \\
 7347 &\dots&\dots&\dots&\dots&6.78 &\dots&6.93 &6.82 &+0.4 &\dots&\dots&\dots&-0.08&\dots& 0.32& 0.21 \\
 7540 &\dots&\dots&\dots&\dots&6.67 &\dots&7.08 &6.93 &+0.4 &\dots&\dots&\dots&-0.04&\dots& 0.47& 0.32 \\
 7912 &\dots&\dots&\dots&\dots&6.75 &\dots&7.01 &6.88 &+0.4 &\dots&\dots&\dots&-0.01&\dots& 0.40& 0.27 \\
 8395 &\dots&\dots&\dots&\dots&6.60 &\dots&6.56 &6.44 &+0.4 &\dots&\dots&\dots&-0.11&\dots&-0.05&-0.17 \\
 8481 &\dots&\dots&\dots&\dots&\dots&\dots&\dots&\dots&+0.4 &\dots&\dots&\dots&\dots&\dots&\dots&\dots \\
 9462 &\dots&\dots&\dots&\dots&6.60 &\dots&6.88 &6.76 &+0.4 &\dots&\dots&\dots&-0.11&\dots& 0.27& 0.15 \\
 9786 &\dots&\dots&\dots&\dots&6.63 &\dots&6.91 &6.79 &+0.4 &\dots&\dots&\dots&-0.04&\dots& 0.30& 0.18 \\
10695 &\dots&\dots&\dots&\dots&6.62 &\dots&6.76 &6.64 &+0.4 &\dots&\dots&\dots&-0.09&\dots& 0.15& 0.03 \\
10806 &\dots&\dots&\dots&\dots&6.62 &\dots&6.81 &6.68 &+0.4 &\dots&\dots&\dots&-0.09&\dots& 0.20& 0.07 \\
10898 &\dots&\dots&\dots&\dots&6.59 &\dots&6.76 &6.59 &+0.4 &\dots&\dots&\dots&-0.15&\dots& 0.15&-0.02 \\
11814 &\dots&\dots&\dots&\dots&6.54 &\dots&6.81 &6.61 &+0.4 &\dots&\dots&\dots&-0.19&\dots& 0.20& 0.00 \\
11938 &\dots&\dots&\dots&\dots&6.69 &\dots&6.78 &6.66 &+0.4 &\dots&\dots&\dots&-0.02&\dots& 0.17& 0.05 \\
12333 &\dots&\dots&\dots&\dots&6.72 &\dots&6.78 &6.66 &+0.4 &\dots&\dots&\dots&-0.02&\dots& 0.17& 0.05 \\
12823 &\dots&\dots&\dots&\dots&6.74 &\dots&7.23 &6.94 &+0.4 &\dots&\dots&\dots& 0.11&\dots& 0.62& 0.50 \\
\hline
\end{tabular}
\label{tabonline}
\tablecomments{Solar values adopted here are: 
 Fe=7.54, Ca=6.27,  O=8.79,   Ni=6.28,
Na=6.21 \citep{gra03}; C=8.52, N=7.92 \citep{gs98}}
\label{tab-abund}
\end{table*}

\end{document}